\documentclass[12pt]{article}
\usepackage[]{alpha}

\usepackage{amsmath}
\usepackage{tabularx}
\usepackage{booktabs}
\usepackage[]{appendix}
\usepackage[utf8]{inputenc}

\newcommand\ev[1]{\left\langle#1\right\rangle}

\begin{document}

\title{A multilevel algorithm for flow observables in gauge theories}

\author[DESY,HU]{Miguel~Garc\'ia~Vera}
\author[DESY]{Stefan~Schaefer}

\address[DESY]{John von Neumann Institute for Computing (NIC), DESY, \\
    Platanenallee 6, D-15738 Zeuthen, Germany }
\address[HU]{Insitut f{\"u}r Physik, Humboldt Universit{\"a}t zu Berlin, \\
    Newtonstr. 15, D-12489 Berlin, Germany}
    
\preprintno{DESY 16-016}

\begin{abstract}
We study the possibility of using  multilevel algorithms for
the computation of correlation functions of gradient flow observables.
For each point in the correlation function an approximate flow is defined
which depends only on links in a subset of the lattice. Together with 
a local action this allows for independent updates and consequently a
convergence of the Monte Carlo process faster than the inverse square root of the number
of measurements. We demonstrate the feasibility of this idea in the correlation
functions of the topological charge and the energy density.
\end{abstract}

\maketitle

\section{Introduction}\label{sec:Introduction}

In Monte Carlo simulations of Yang-Mills gauge theories, correlation functions 
of gluonic operators suffer from a severe signal-to-noise problem~\cite{Parisi:1983ae}. 
While the signal of a two-point function itself falls off exponentially with 
the distance between the operators, the variance is largely independent of 
their separation. Since the error decreases only like $1/\sqrt{N}$, with 
$N$ the number of measurements, this makes their measurement in numerical calculations 
for large separations exceedingly difficult.

Due to their favourable renormalization properties, correlation functions of observables 
defined through the Yang-Mills gradient flow are an important tool 
to study gauge theories~\cite{Narayanan:2006rf,Luscher:2010iy,Luscher:2011bx}. 
In particular, they allow for a computationally economical definition of 
the topological susceptibility on the lattice
\begin{equation*}
\chi_\mathrm{top}=\frac{1}{V}\int  dx \, dy \, \langle q(x) q(y) \rangle \,,
\end{equation*}
with the topological charge density $q(x)$ defined through the Wilson flow.

The signal-to-noise problem present at large distances in the $\langle
q(x)q(y)\rangle$ correlation function translates into a lack of volume
averaging of $\chi_\mathrm{top}$: the statistical error of the susceptibility
from a given number of configurations does not improve with increasing volume.
In pure gauge theory this can be partially overcome with large statistics, but in practice, 
rather small lattices are still used and the finite size effect needs to
be carefully controlled~\cite{Ce:2015qha}. In large volume, it is therefore 
beneficial to study the dependence of $\langle q(x) q(y) \rangle$  on $|x-y|$ 
directly and model the large distance behavior~\cite{Bazavov:2010xr} or integrate 
it such that the contribution of the tail can be neglected~\cite{Bruno:2014ova} 
given the statistical accuracy. 

On a related note, we also point out that it has been suggested to extract the masses of glueballs from
the large distance behavior of two-point functions of the  smoothed topological 
charge and energy density~\cite{Chowdhury:2014kfa}. For this 
approach to work, it is highly beneficial to have a precise determination of the 
tail of the correlator at large distances.

One way to deal with the signal-to-noise problem is to use multi-level algorithms 
which rely on the locality of the observable and of the action~\cite{Parisi:1983hm,Luscher:2001up}.
If it is possible to decompose the observables in contributions from different
parts of the lattice, each of them can be updated independently.  Depending on the number
of sub-lattices and the efficiency of the decomposition, the signal-to-noise problem
can be eliminated or at least reduced substantially. Recently, such type of algorithms 
have also been adapted to the case of quenched lattice QCD to compute fermionic 
correlators~\cite{Ce:2016idq}. 

In the case of flow observables, multi-level algorithms can not be applied directly 
basically due to the fact that the flow has a footprint which is not finite. In this 
paper we propose a first step in the direction of solving this problem. We study a two-level 
algorithm where the lattice is decomposed into two sub-volumes and observables 
are defined such that they depend only on the fields in the respective sub-volume.

The paper is organized as follows. In Sect.~\ref{sec:Algorithm} we describe the 
algorithm and in Sect.~\ref{sec:NumericalObs} we define the observables that 
we use in this study. Then, in Sect.~\ref{sec:Results} we demonstrate the 
feasibility of this setup and discuss how the improvement works before we conclude.

\section{Algorithm}\label{sec:Algorithm}

In order to make the discussion of the algorithm as self contained as possible,
we shall briefly present the main ideas introduced in~\cite{Luscher:2001up,Meyer:2002cd}, 
in a context which is directly applicable to our case.

\subsection{Factorized observables}

For simplicity, we consider $\mathrm{SU}(N)$ Yang-Mills gauge theory on the lattice with the standard Wilson
action, although more general type of actions can be used,

\begin{equation}\label{eq:Action} S\left[ U \right] = \frac{\beta}{N} \sum_P
\text{Tr} \left\lbrace 1 - U \left( P \right) \right\rbrace \,, \end{equation}
where $U(P)$ is the product of gauge links around the plaquette $P$. 

\begin{figure}\centering \includegraphics[width=0.5 \textwidth]{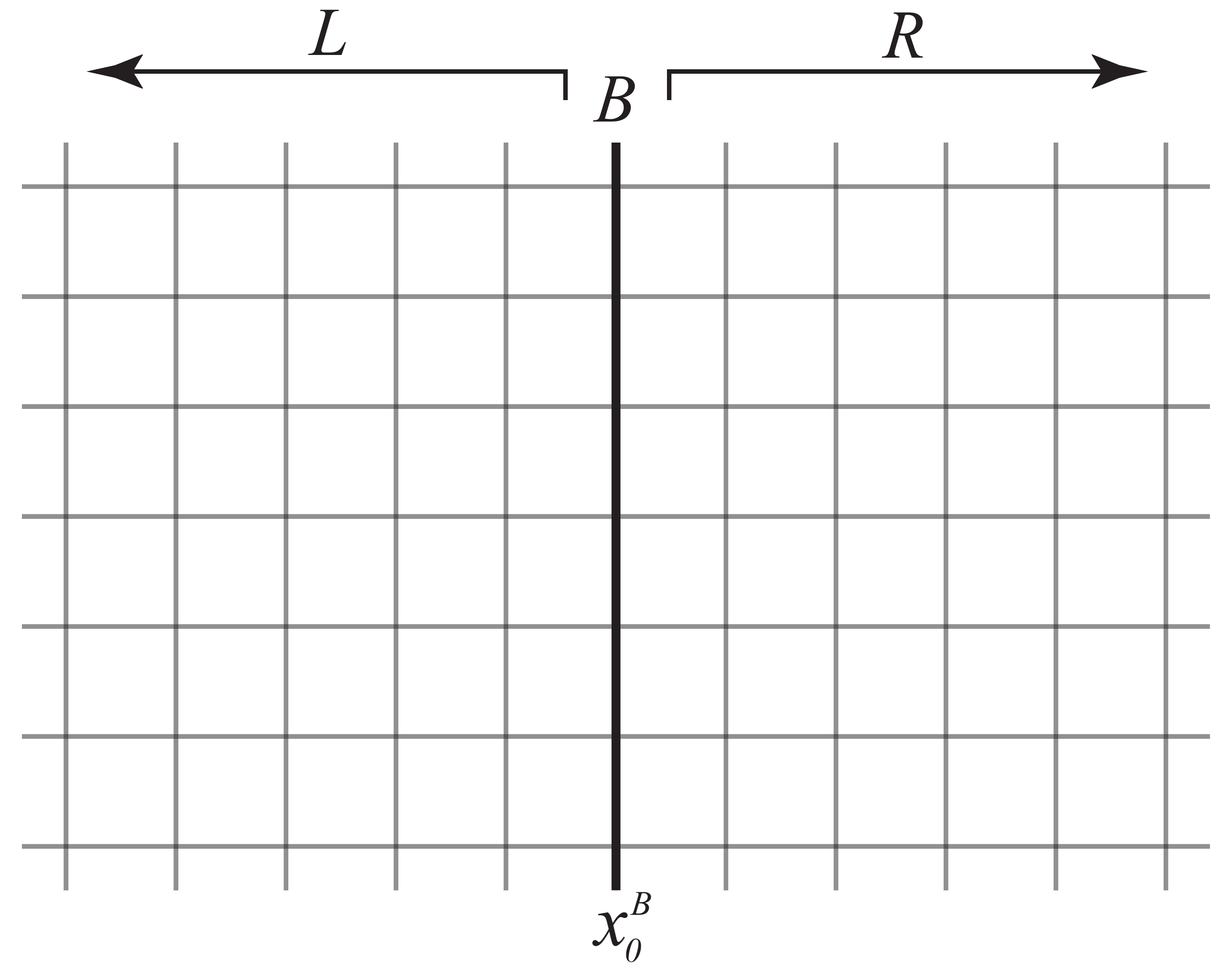}
\caption{Factorized lattice setup. The lattice is split into two sub-volumes 
$L$ and $R$ which are separated by the boundary links $B$ defined as the spatial 
links at the timeslice $x_0^B$.}\label{fig:Lattice_setup} \end{figure}

Take $B$, $L$ and $R$ to be three disjoint subsets of gauge links, such that 
they make up for the totality of gauge links on the lattice. We choose 
$B$ in such a way that the gauge action $S\left[U_L,U_B,U_R\right]$ can be 
decomposed as $S_L\left[U_L,U_B\right] + S_R\left[U_B,U_R\right] + S_B\left[U_B\right]$, 
where by $U_{L,B,R}$ we refer to the set of gauge links which belong to $L$, $B$ 
and $R$ respectively. One natural choice for $B$ is the subset 
of all spatial links at a fixed time-slice $x_0^B$, so that, $L$ and $R$ are simply 
defined as all gauge links that are located to the left or to the right of the boundary $B$. 
This setup is depicted in Fig.~\ref{fig:Lattice_setup}.

For two observables $\mathcal{O}(x)$, and $\mathcal{O'}(y)$, which are defined  
for $x \in L$ and $y \in R$, the decomposition of the action makes it 
possible to write

\begin{align}\label{eq:fact_obs} \left\langle \mathcal{O} \mathcal{O'}
\right\rangle &= \int dU_B \, p(B) \, \left[ \mathcal{O} \right]_{L}
\left[ \mathcal{O'} \right]_{R} \, ,\\\nonumber
 \left[ \mathcal{A}\right]_{L,R} &= \frac{1}{Z_{L,R}}\int dU_{L,R} \, \mathcal{A} \, e^{-S_{L,R}[U_{L,R},U_B]}
\end{align}
where $\mathcal{A}$ is either $\mathcal{O}$ or $\mathcal{O'}$, $Z_L$ and $Z_R$ 
are the normalization factors such that $\left[1\right]_{L,R}=1$,  
and $p(B) = \frac{Z_L Z_R}{Z} e^{-S[U_B]}$, with $Z$ the standard partition function.

Eq. \eqref{eq:fact_obs} expresses the fact that one can average an observable over $L$
and $R$ independently while keeping $B$ fixed and then take the average over
the possible values of $B$. As discussed in~\cite{Luscher:2001up,Meyer:2002cd}, 
this process can be iterated if the
operators $\mathcal{O}$ or $\mathcal{O'}$ can be subsequently factorized.
This is the property of factorization that was exploited originally in
\cite{Luscher:2001up} to show an exponential reduction in the error of the
expectation value of large Wilson loops.

The idea presented above can be realized in a Monte Carlo simulation as follows. 
First generate $N_0$ regular updates which are used to perform the integration 
over $U_B$ in Eq. \eqref{eq:fact_obs}. Then, for each of the $N_0$ 
original configurations, $N_1$ updates of $L$ and $R$ are done independently 
while keeping $B$ fixed, so that for the product 
$\left[\mathcal{O}\right]_L \left[\mathcal{O'}\right]_R$, the error decreases 
ideally as $1/N_1$ instead of the standard $1/\sqrt{N_1}$. As shown in the Appendix~\ref{sec:Appendix} 
this can be reached only for operators with vanishing expectation value $\ev{\mathcal{O}} = \ev{\mathcal{O'}} = 0$. Therefore, in the following we restrict ourselves to the connected correlation functions.

Note that factorization makes it possible to obtain a better 
scaling for the errors in $\left[\mathcal{O}\right]_L \left[\mathcal{O'}\right]_R$, but the 
error on the final expectation value $\ev{\mathcal{O}\mathcal{O'}}$ depends on the average over $B$ which 
scales as $1/\sqrt{N_0}$. This means that for large values of $N_1$ the error is 
controlled by the fluctuations of $B$ and hence the dominant scaling will be the $1/\sqrt{N_0}$.
As discussed in the following sections and as shown in the Appendix~\ref{sec:Appendix}, 
in practice one can take very large values of $N_1$ before the ideal scaling is no 
longer valid.

\subsection{Modified flow Observables}

Given the gauge link variables $U(x,\mu)$, the flow variables $V^t(x,\mu)$
associated to them are defined by the equation
\cite{Luscher:2009eq,Luscher:2010iy,Narayanan:2006rf}

\begin{equation}\label{eq:Wflow} \dot{V}^t(x,\mu)= -g_0^2 \left\lbrace
\partial_{x,\mu} S(V) \right\rbrace V^t(x,\mu), \qquad V^{t=0}(x,\mu) = U(x,\mu)
\end{equation}

The effect of the flow can be viewed as a smoothing of the gauge fields over a
spherical range with a mean square radius of $\sqrt{8t}$. Because of this, any observable 
defined in $L$ or $R$ has a non-trivial dependence on gauge links from the opposite 
domain at positive flow time $t$, and it can not be factorized as 
required for Eq. \eqref{eq:fact_obs} to hold. However, the smoothing produced 
by the flow is exponentially suppressed at large distances, which leads us to propose a slightly 
modified version of the flow equations, such that an observable computed with the modified 
flow gauge links $\tilde{V}^t$ is a good approximation to the original one and 
can be factorized as required in Eq.~\eqref{eq:fact_obs}.

If the Wilson action is also used in the definition of the flow, we propose the
following modified flow equation

\begin{equation}\label{eq:Wflow_modified} \dot{\tilde{V}}^t(x,\mu)=
\begin{cases} -g_0^2 \left\lbrace \partial_{x,\mu} S(\tilde{V}^t) \right\rbrace \tilde{V}^t(x,\mu), \quad \tilde{V}^{t=0}(x,\mu) = U(x,\mu), & \text{if } U(x,\mu) \in L\cup R.\\ U(x,\mu), & \text{if } U(x,\mu) \in B. \end{cases} \end{equation}

The modified version accounts for integrating the flow equations while the 
links at $B$ are kept fixed. It is constructed such that for each link $U(x,\mu) \in L$, also 
the smoothed link $\tilde{V}^t(x,\mu)$ only depends on links in $L$ and $B$. 
Therefore, for an observable $\mathcal{O}(x)$, in either $L$ 
or $R$, the modified flow observable $\widetilde{\mathcal{O}}^t(x)$ does not get any contribution 
from the links in the opposite domain. If $\widetilde{\mathcal{O}}^t(x)$ 
is a good approximation of $\mathcal{O}^t(x)$, one can take advantage of factorization 
to obtain a better scaling of the errors of $\ev{\mathcal{O}\mathcal{O'}}$ with 
respect to the $N_1$ nested Monte Carlo updates.

\subsection{Two point correlation function}

We now consider the case of the connected two point correlation function 
$\mathcal{O}(x)\mathcal{O}(y)$ for $x$ and $y$ spacetime points in the four 
dimensional lattice, and put together the modified flow observables with the 
multi-level scheme. We define

\begin{equation}\label{eq:correlator}
C_{\mathcal{O}}^t(x,y) = \ev{\mathcal{O}^t(x) \mathcal{O}^t(y)}_C = 
\ev{\mathcal{O}^t(x) \mathcal{O}^t(y)} - \ev{\mathcal{O}^t(x)}\ev{\mathcal{O}^t(y)}\, ,
\end{equation}
as the connected ($C$) correlation function of the observable $\mathcal{O}$. 
If the two points $x$ and $y$ are separated from the boundary $B$ by a distance 
much larger than the radius of the flow $\sqrt{8t}$, then the modified 
version of Eq.~\eqref{eq:correlator} using the gauge links $\tilde{V}^t$ 
is a good approximation to the original correlator. To show this, we look at the correction term 
$\Delta$, defined as the difference between the flow observable and the 
observable computed using the modified flow

\begin{equation}
\label{eq:corr_term} 
\Delta_{\mathcal{O}}^t(x,y) = C^t_{\mathcal{O}}(x,y) - \widetilde{C}^t_{\mathcal{O}}(x,y) \, .
\end{equation}

Notice that we have left the dependence on both $x$ and $y$ explicit, as the presence of the 
boundary $B$ breaks full translation invariance and one must keep track not only 
of the distance $|x-y|$ between source and sink, but also of the distance of 
both $x$ and $y$ with respect to $B$. The reason for this will become evident in the 
next section when we discuss a practical application of the algorithm. 
When not explicitly needed we will drop the $t$ index in every quantity.

For the observables discussed in the next section, our data shows that 
for a sufficiently large separation from $B$ compared to the smoothing radius, $\Delta$ becomes
negligible. In spite of that, our strategy is not to
neglect the correction term $\Delta$. Instead, in a nested Monte Carlo simulation, 
the idea is  to use first the $N_0$ generated configurations to estimate $\Delta$ 
and then use this estimation to correct for the value of $\widetilde{C}_{\mathcal{O}}(x,y)$. 
For this to work, we need that the fluctuations of $\Delta$ are much 
smaller than the fluctuations of $C_{\mathcal{O}}$ in such a way that 
we can use the $N_0$ updates to estimate $\Delta$ and subsequently perform 
the $N_1$ nested Monte Carlo updates independently in $L$ and $R$ 
to compute $\widetilde{C}_{\mathcal{O}}$.

The main equation of this paper is a modified version of Eq.~\eqref{eq:fact_obs} 
which takes into account the correction term $\Delta$ and is applicable 
for any two point correlation function of Wilson flow observables. 
We define an estimator $\widehat{C}_{\mathcal{O}}^t(x,y)$ of 
$C^t_{\mathcal{O}}(x,y)$ as

\begin{align}\label{eq:main} \widehat{C}_{\mathcal{O}}^t(x,y) &= \frac{1}{N_0}\sum_{N_0}
\left\lbrace \left[\widetilde{\mathcal{O}}^t(x)\right]_L \left[\widetilde{\mathcal{O}}^t(y)\right]_R +
\Delta^t_{\mathcal{O}}(x,y) \right\rbrace\\\nonumber
\left[\widetilde{\mathcal{O}}^t(z)\right]_{L,R} &= \frac{1}{N_1} \sum_{N_1}
\widetilde{\mathcal{O}}^t(z) \, , \qquad z=x,y \, , \end{align}
where $(x,y) \in L \times R$. The estimator in Eq.~\eqref{eq:main} is correct up to 
errors of order $O(1/\sqrt{N_0})$, which comes from the fact that $\Delta$ is only 
computed on the $N_0$ standard updates. However, in the next section we show that 
the fluctuations of $\Delta$ are exponentially suppressed with the distance 
to the boundary, so that the leading term for the scaling of the error in $\widehat{C}$ 
comes from the correlator of the modified flow observables.

\section{Numerical test of the modified flow observables}\label{sec:NumericalObs}

To test our algorithm we work with the $\mathrm{SU}(3)$ gauge group and a set of gauge
configurations generated with the parameters shown in Table
\ref{tab:Ensembles}. The configurations are generated for a value of $\beta = 6.11$, which 
corresponds to a lattice spacing of $a \approx 0.08 \, \text{fm}$ and a effective 
smearing radius $\sqrt{8 t_0} \approx 6 a$. Open boundary conditions are used in the time direction
\cite{Luscher:2011kk}. We consider two observables, the topological charge density
$q$ and the Yang-Mills energy density $e$. In particular, we look at the
connected two point correlation function of the timeslice summed 
$\bar{q}$ and $\bar{e}$

\begin{table} 
\centering 
\begin{tabular}{cccccc} 
\toprule 
$\beta$   & $(T/a) \times (L/a)^3$    & $t_0/a^2$     & $a \, \text{[fm]}$ & $N_0$\\
\midrule $6.11 $  & $80 \times 20^3$ & $4.5776(15)$   &  $0.078$ & $384$\\ 
\bottomrule 
\end{tabular} 
\caption{Lattice
parameters. We report the lattice bare coupling $\beta$, the lattice dimensions $L$ and 
$T$, the scale parameter $t_0$ defined in~\cite{Luscher:2010iy}, the lattice spacing 
$a$ computed using the $r_0 = 0.5 \, \text{[fm]}$ scale from \cite{Necco:2001xg}, and the number of 
generated configurations $N_0$.}
\label{tab:Ensembles} \end{table}

\begin{align} 
C_q^t(x_0,r) &= \ev{\bar{q}^t(x_0) \bar{q}^t(x_0+r)}_C, 
\qquad
\bar{q}^t(x_0) = a^3\sum_{\vec{x}} q^t(\vec{x},x_0)\\
\nonumber C_e^t(x_0,r) &=\ev{\bar{e}^t(x_0) \bar{e}^t(x_0+r)}_C, 
\qquad \bar{e}^t(x_0) =a^3 \sum_{\vec{x}}e^t(\vec{x},x_0) \, ,
\end{align}
where we have left the $x_0$ dependence explicit in order to keep track of the
distance to the boundary $B$, which is chosen to be the subset of spatial links with 
time coordinate $x^B_0=T/2$. All computations are done in such a way that both $x_0$ 
and $x_0 + r$ are placed far enough from the open boundaries. From now on we 
shall use $\mathcal{O}$ to refer to either $q$ or $e$ when there is no need to make a 
distinction between them.

We use the 384 independent configurations to study the dependence of the fluctuations 
of $\Delta$ on both $x_0$ and $r$. First we consider the correlators which are
symmetric with respect to $B$, so we choose a source which is placed at the value 
of $x_0 = (T -r)/2$. In this case, the correlator is only a function of 
$r$ and given by $C_{\mathcal{O}}(r) = C_{\mathcal{O}}\left(\left(T-r\right)/2,r\right)$.

\begin{figure} \includegraphics[width=\textwidth]{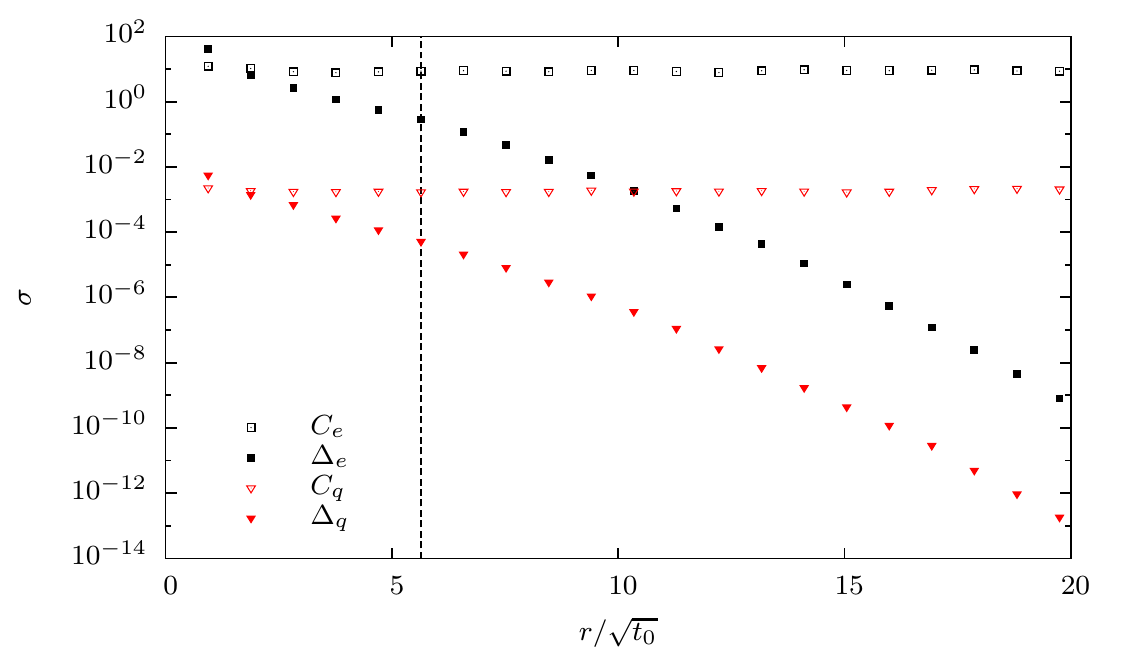}
\caption{Statistical error $\sigma$ of $\Delta_{\mathcal{O}}$ and $C_{\mathcal{O}}$ at flow time
$t=t_0$. For both observables, $\bar{e}$ and $\bar{q}$, the error in the correlator is independent 
of the distance $r$, but the errors of $\Delta$ seem to decay at least exponentially with the 
distance. The dotted vertical line is added as a reference to indicate the point where the distance from the boundary is equal to the smoothing radius $\sqrt{8t_0}$. For larger separations, the errors in $\Delta_{\mathcal{O}}$ are below 5\% those of $C_{\mathcal{O}}$. Uncertainties are smaller than the data markers.}\label{fig:Errors_Delta_r} \end{figure}

Fig.~\ref{fig:Errors_Delta_r} shows the dependence of the error of both
$C_{\mathcal{O}}(r)$ and $\Delta_{\mathcal{O}}(r)$ for a 
fixed value of the flow time $t=t_0$. The errors are computed by measuring the autocorrelation 
function as described in \cite{Wolff:2003sm}. We note that for separations from the boundary 
larger than the smoothing radius, the fluctuations in $\Delta$ are below 5\% of those
of the observable. As will be discussed in Sect.~\ref{sec:Parameters_choice}, 
the fact that the ratio between the fluctuations of the observable and those of the 
correction term decrease at large distances contributes to the fact that the 
algorithm is efficient up to very large values of $N_1$.

Since the effective smearing radius produced by the flow grows as $\sqrt{t}$, the
effect of freezing the boundary links at $B$ increases monotonically with the
flow time. We have observed this behaviour in our data, but we are more 
interested in the behaviour of the correlation functions at the reference flow scale 
$t =t_0$. For different values of the flow, a similar analysis can be performed. 
However, it is clear that if the fluctuations of $\Delta$ are ``small'' for a given 
value of $t'$, they are also small for $t < t'$.

\begin{figure} \includegraphics[width=\textwidth]{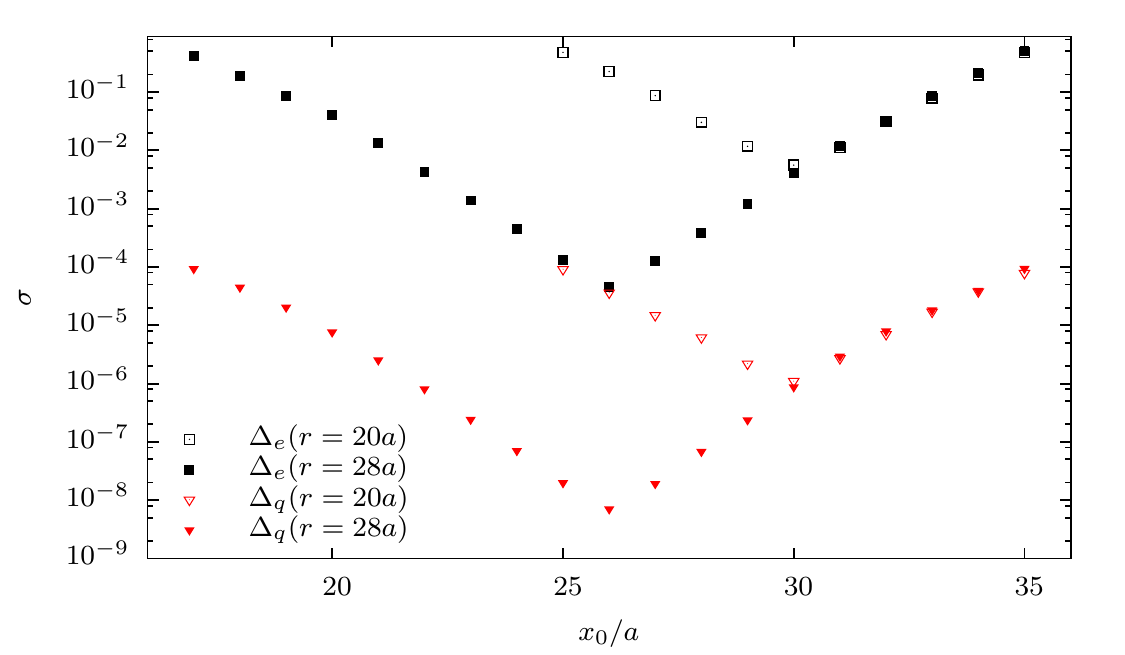} 
\caption{Statistical error $\sigma$ of $\Delta_{\mathcal{O}}$ as a function of $x_0$ 
for two values of $r$ at flow time $t=t_0$. Open symbols correspond to a value of 
$r=20 a = 9.4 \sqrt{t_0}$, while filled symbols to a value of $r = 28 a = 13.2 \sqrt{t_0}$. The smallest error corresponds 
to the symmetric point in which both source and sink are placed far from the boundary. 
Uncertainties are smaller than the data markers.}\label{fig:Errors_Delta_x0}\end{figure}

Next, we go beyond the symmetric case and look at the $x_0$ dependence of $\Delta$.
Fig.~\ref{fig:Errors_Delta_x0} shows a plot of the errors in $\Delta$ as a function of
$x_0$ for two fixed values of $r$ at $t=t_0$. 

Notice that the effect of the flow is that of a Gaussian smearing, so we should 
expect that the errors in $\Delta$ decay at least exponentially with the distance 
to the boundary $B$. Both Figs.~\ref{fig:Errors_Delta_r} 
and~\ref{fig:Errors_Delta_x0} show a behaviour which is compatible with this statement.

The results presented in this section show that using the modified flow
equations has little impact in the two point function, and the effect can be
incorporated in the correction term $\Delta$. When using equation
\eqref{eq:main} it is important to tune the value of $N_0$ and $N_1$ in such a
way that the effect of $\Delta$ remains under control. In particular, due to
the exponential smoothing of the flow, $N_1$ can be chosen larger at larger
values of $r$, which is precisely where a higher precision is required.

\section{Results}\label{sec:Results}

We consider the ensemble in Table \ref{tab:Ensembles} and for each of the $N_0$
configurations we perform $N_1 = 40$ Monte Carlo updates while keeping $B$
fixed. The updates are separated by 60 sweeps, where one sweep is composed of 
8 over-relaxation updates followed by 1 heat-bath update. 
Both updates are performed using the Cabibbo-Marinari technique 
applied to three $\mathrm{SU}(2)$ subgroups~\cite{Cabibbo:1982zn,Brown:1987rra}.

In the following we present our findings concerning the scaling of the 
errors with respect to $N_1$ and show the application of our algorithm 
for the computation of the two point correlation function over the whole range 
of distances allowed in our finite size lattice. The limitations of the method 
are also discussed. We conclude this section by using our method to compute the 
topological susceptibility and compare it to the result obtained with the standard 
algorithm.

\subsection{Autocorrelation times}

An interesting question to explore is whether or not an undesirable growth of 
the autocorrelations is introduced due to the freezing of the boundary $B$. Such an 
effect could have an impact on the cost of the measurement in our nested Monte Carlo 
scheme. To investigate that, we look at the integrated autocorrelation 
time $\tau_{\mathrm{int}}$ of $\mathcal{O}(x_0)$ as a function of the time coordinate $x_0$.
Given that the $N_0$ standard updates are completely decorrelated, the relevant 
autocorrelation function is given by the average over $N_0$ of the autocorrelation 
function for the $N_1$ nested updates,
$\widetilde{\Gamma}(t) = \frac{1}{N_0} \sum_{i=1}^{N_0} \Gamma^i(t)$. Where $\Gamma^i(t)$ is 
precisely the autocorrelation function for each of the nested chains. 

Now, $\tau_{\mathrm{int}}$ can be defined in the usual way~\cite{Wolff:2003sm} in 
terms of the average autocorrelation function $\widetilde{\Gamma}(t)$. Our data 
shows that $\tau_{\mathrm{int}}$ increases at most by a factor of $1.5$ when 
the observables approach the boundary $B$, so there is not a significant effect. 
However, on different observables, it could have a more severe impact which 
then must be taken into account when spacing the $N_1$ nested updates and 
calculating the cost of the simulation.

\subsection{Choice of the parameters}\label{sec:Parameters_choice}

The introduction of the nested updates adds an extra parameter to be 
tuned in the algorithm, as the parameter $N_1$ can be chosen to minimize 
the errors at a given computational effort. We argue that for the connected 
correlator $\widehat{C}_{\mathcal{O}}$, when source and sink are placed far away 
from the boundary, the value of $N_1$ up to which the algorithm is efficient can 
be scaled exponentially with respect to the distance to $B$. 

To show this, in appendix~\ref{sec:Appendix} we 
have looked into the scaling of errors with respect to $N_0$ and $N_1$ in a 
Monte Carlo simulation. Our results show that the leading contribution to the 
error in the connected correlator scales as $1/\sqrt{N_0}N_1$, which corresponds 
to the ideal case, but additionally there are other terms that scale as $1/\sqrt{N_0N_1}$ and as 
$1/\sqrt{N_0}$. Such terms however; when dealing with connected correlation functions, 
are exponentially suppressed as $e^{-m_0|x^M_0 - x^B_0|}$, 
where $x^M_0$ is the time coordinate of either source or sink, whichever is the closest to 
the boundary, and $m_0$ is the mass of the lightest mode which is compatible with 
the symmetries of $\mathcal{O}$. This means that we can expect the ideal scaling up to very large 
values of $N_1$ given that source and sink are far away from the boundary in units of $1/m_0$.

Another effect that must be taken into account is the presence of the correction 
term $\Delta$. Such term is measured only over the $N_0$ standard updates, so that 
its error should scale in the standard way as $1/\sqrt{N_0}$. This will add another 
term which is independent of $N_1$ to the final error. We can see from  
our results in Fig.~\ref{fig:Errors_Delta_x0} that for a fixed $N_0$, the error in $\Delta$ decays 
at least exponentially fast with the distance of either source or sink to the boundary 
$B$. This means that for the final estimator $\widehat{C}$, the value of $N_1$ up to which the ideal 
scaling is valid increases exponentially with the distance to the boundary as long as 
$|x^M_0 - x^B_0|$ is larger than the relevant scale, either $1/m_0$ for the effects coming 
from $\widetilde{C}$ or $\sqrt{8t}$ for those coming from $\Delta$.

\subsection{$N_1$ dependence of the error}\label{sec:Scaling_errors}

\begin{figure} \includegraphics[width=\textwidth]{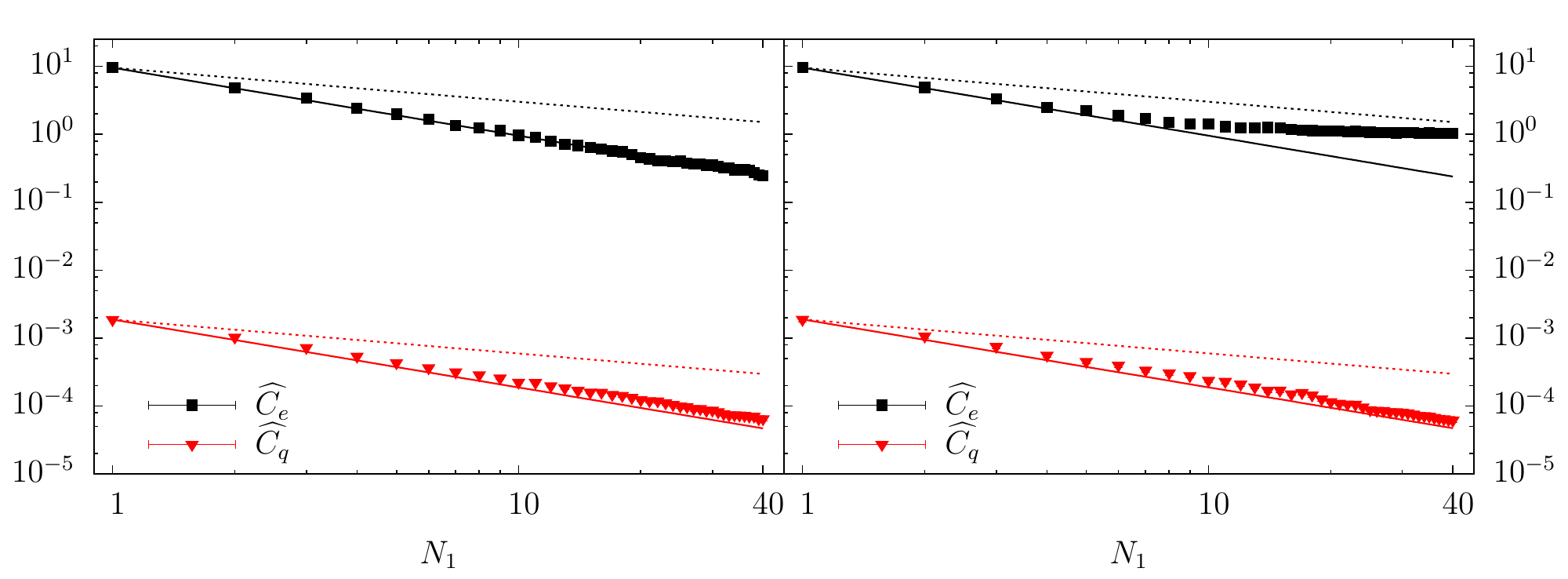}
\caption{Scaling of the error of $\widehat{C}_{\mathcal{O}}$ as a function of 
$N_1$. On the left for a value of $r= 28 a = 13.2 \sqrt{t_0}$ and on the right for a 
value of $r=14 a = 6.6 \sqrt{t_0}$. The solid line indicates a scaling of the 
error proportional to $1/N_1$, while the dotted line corresponds to the standard 
$1/\sqrt{N_1}$ scaling. For the smaller value of $r$ (right plot), 
we observe a saturation in the number of effective $N_1$ nested updates that 
can be used to reduce the errors. In fact, after $N_1 \approx 6$ we observe no 
significant improvement. }\label{fig:Scaling_errors} \end{figure}

To show the way in which our algorithm improves over the standard one, we
measure the scaling of errors with respect to $N_1$ for the symmetric
correlator. The results for two different values of $r$ are shown in Fig.~\ref{fig:Scaling_errors}. 
For the larger $r = 28 a$, and for $N_1=40$, we 
are still in the regime where the ideal scaling is the dominant one, so on 
the left subplot we see a scaling of the error which is compatible with 
$1/N_1$ for the whole range of $N_1$ values.

For the smaller value of $r = 14 a$, in particular when looking at the case 
of $\widehat{C}_e$, we observe that for $N_1 \gtrsim 6$ the error improves only 
marginally with $N_1$, which means that we are already in the regime where the 
term independent of $N_1$ becomes relevant. This supports the discussion of the 
previous section and shows that for small values of $r$ 
there is no significant improvement by performing a very large number of $N_1$ nested Monte Carlo 
updates. In practice, one can use all the $N_1$ generated nested updates for all 
values of $r$, but for small separations, the effect of using all of them 
is not significative. 

For a given $N_0$ and $N_1$, the value of $r$ at which the ideal scaling is not 
valid anymore is observable dependent and so it has to be studied on a case by case basis. 
In our particular case, we observe that for $N_1 =40$ we are on the ideal scaling 
regime for the correlator at distances starting at values of 
$r = 16 a = 7.5 \sqrt{t_0}$ at a flow time $t=t_0$. 

\subsection{Application of the algorithm}\label{sec:Application_algo}

To show how the algorithm performs for the whole range of distances in the 
two point correlator, we compute $C_q$ and $C_e$ using the standard
algorithm and using our nested Monte Carlo scheme. 
For each value of $r$ and $x_0$ we compute $C_{\mathcal{O}}$ and $\widehat{C}_{\mathcal{O}}$. 
We use the $N_0=384$ standard updates to compute $C_{\mathcal{O}}$ in the usual way. For 
our nested algorithm we employ the $N_1=40$ nested updates for each of the standard ones. 

When using the standard approach, the correlator at distance $r$ is computed by averaging 
over all the $x_0$ values in the plateau region. In the case of our algorithm this is  not the best strategy, 
as translation invariance is lost due to the presence of the boundary $B$. Instead, we find 
it beneficial not to use those timeslices for which source or sink are closer to $B$ than 
a given distance $r_B$, which is tuned as part of the analysis. When 
working at $t=t_0$ we find the best choice to be $r_B = 6a$, which 
is compatible with the smearing radius $\sqrt{8t_0} \approx 6a$.

The inclusion of $r_B$ in the analysis means that for separations smaller 
than $2 r_B$ the average is done only when source and sink are in the same 
domain, either $L$ or $R$. In those cases, we expect no improvement with respect 
to the standard algorithm. For larger distances however, one can choose to have 
$x_0 \in L$ and $x_0 +r \in R$, where the better scaling is expected. Notice that 
for intermediate distances, the average over timeslices would also include terms 
for which source and sink are in the same domain. These terms would contribute to 
the error with the usual scaling $1/\sqrt{N_0N_1}$, so we find the better performance 
when they are also not included in the average and we sum only over the factorized 
terms.

We also look at smaller values of the flow time $t$, in particular we look at a value of 
$t=t_0/10$. Smaller flow times can be of interest if one is looking at obtaining 
the glueball masses. In such cases, the analysis is the 
same as described above, but only the value of $r_B$ changes; for example, at 
$t=t_0/10$ we find an optimal value of $r_{B}=3 a$, which is also compatible with 
the value of the smearing radius.

\subsection{Performance of the algorithm}

\begin{figure} \includegraphics[width=\textwidth]{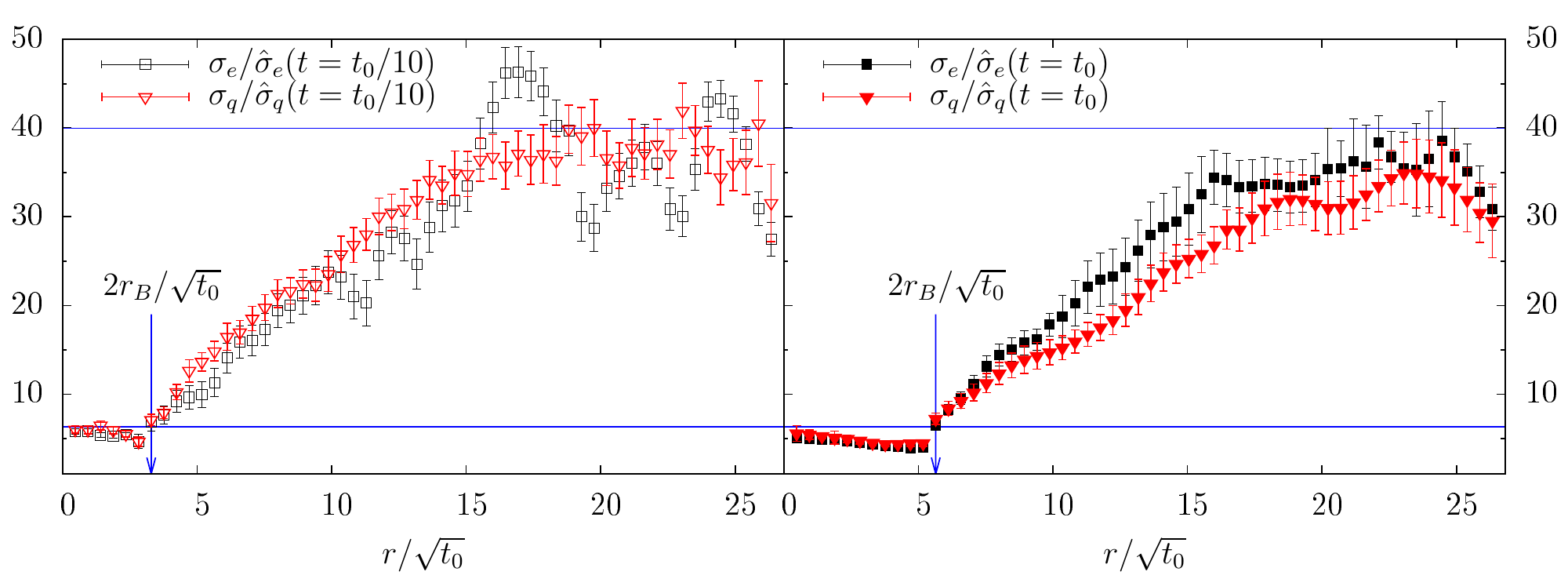}
\caption{Ratio of the errors $\sigma_{\mathcal{O}}/\hat{\sigma}_{\mathcal{O}}$ 
as a function of $r$. Open symbols are the results for a flow time $t=t_0/10$, while 
filled symbols corresponds to the value of $t=t_0$. One can see that the improvement 
can be split into three distinct regions. For short distances, our algorithm is not as 
efficient as the standard one. For intermediate distances our algorithm is already better 
than the standard one but does not reach the theoretical maximum improvement, which is 
only achieved in the large distance regime. The two horizontal lines represent the 
theoretical maximum improvement of the standard algorithm and the one expected from our algorithm. 
}\label{fig:ErrorRatio} \end{figure}

We apply the strategy described above to compute the $\widehat{C}_{\mathcal{O}}$ 
and $C_{\mathcal{O}}$ correlators for a wide range of separations $r$ between 
source and sink. To assess the performance of the algorithm, in Fig.~\ref{fig:ErrorRatio} 
we plot the ratio between the error of the standard correlator $\sigma_{\mathcal{O}}$ and 
the error of the improved one $\widehat{\sigma}_{\mathcal{O}}=\sigma(\widehat{C}_{\mathcal{O}})$. 
With the standard algorithm, if the statistics are increased by a factor $N_1=40$, the error 
should scale down by a factor $\sqrt{N_1} \approx 6.3$. The lower horizontal line in Fig.~\ref{fig:ErrorRatio} shows the theoretical improvement of the standard algorithm for 
the same statistics as the ones we use in our two-level algorithm.  

For the short distance region, we observe an improvement which is below the theoretical 
one of the standard algorithm. As explained before, this is expected due to the fact 
that one can not make full use of translation invariance and the our algorithm is not 
designed to be the most efficient for such short distances when the effects of the 
flow are more relevant.

As soon as $r \geq 2 r_{B}$ one enters the region where the new algorithm outperforms the standard one. 
This is expected as for most of these values of $r$ we can make full use of the $N_1 = 40$ nested 
updates. However, at intermediate distances, we lose due to the lack of translation invariance in 
the $x_0$ direction. This is precisely what we observe as the improvement rises 
continually from $r=2 r_B$ until it reaches the theoretical maximum improvement 
equal to $N_1=40$. For values of $r$ sufficiently large, our algorithm performs as 
expected and we obtain the theoretical maximum improvement which is shown in 
the figure by the upper horizontal line. We observe the same qualitative behaviour 
for different values of the flow time, the only difference being the different 
value of $r_B$ which is used in the analysis. Clearly, for smaller values 
of the flow time we are able to outperform the standard algorithm at 
even shorter distances, which could be useful for certain applications.

\subsection{Topological susceptibility}

As a final test of our proposal, we compute the topological susceptibility $\chi$ 
at $t=t_0$ and compare it to the result obtained when using the standard algorithm. For the 
comparison we use the same statistics in both cases, i.e, $N_0N_1 = 15360$ measurements, 
so that the computational effort is roughly the same. 
For the definition of the susceptibility we use the one in~\cite{Bruno:2014ova}. To write this 
in terms of our observables, we define $\overline{C}_q(r)$ as the average of $C_q(x_0,r)$ over 
$x_0$. We proceed as described in the previous section, so for the standard algorithm we average 
over all values of $x_0$, while in the case of the new algorithm we use only those values of $x_0$ 
such that source and sink are not closer than $r_B$ to the boundary $B$. Then, we define the 
topological susceptibility as

\begin{equation}\label{eq:chi_corr_def}
 \chi(r_{\mathrm{cut}}) =  \frac{a}{L^3} 
 \sum_{z_0=-r_{\mathrm{cut}}}^{r_{\mathrm{cut}}}  \overline{C}_q(|z_0|) \, ,
\end{equation}
where $r_{\mathrm{cut}}$ should be chosen so that the statistical error in the sum is larger than the 
estimated systematic error from cutting the summation. We are not so interested in 
choosing the best value of $r_{\mathrm{cut}}$ but more on comparing the performance of the 
two-level algorithm with respect to the standard one. 

In Table~\ref{tab:Suscept} we show the results at three different values of $r_{\mathrm{cut}}$ using 
both the standard algorithm and the new nested Monte Carlo algorithm that we propose in 
this paper. As already pointed out in the introduction, with the traditional approach, 
summing up the correlator to large values of $r_{\mathrm{cut}}$ only increases the error 
while the signal remains relatively constant~\cite{Bazavov:2010xr,Bruno:2014ova}. We 
clearly observe this effect in our data when using the standard method. On the other 
hand, the error when using our algorithm remains relatively constant when the 
value of $r_{\mathrm{cut}}$ is increased from values of $0.85 \, \text{fm}$ up to $4.19 \, \text{fm}$. 
In fact, for the largest value of $r_{\mathrm{cut}}$, the improvement when using our algorithm 
is more than twofold, corresponding to an increase in statistics by a factor $5$.

\begin{table}
  \centering
  \begin{tabular}{cccc}
    \toprule
     $r_{\mathrm{cut}}/\sqrt{t_0}$ & $r_{\mathrm{cut}}\, \text{[fm]}$ & Standard &  New \\
    \midrule
     $5.1$  & $0.85$ & $6.405(46)$  & $6.347(60)$  \\ 
     $15.4$ & $2.56$ & $6.507(94)$  & $6.291(61)$ \\
     $25.2$ & $4.19$ & $6.518(164)$ & $6.254(69)$ \\
    \bottomrule
  \end{tabular} 
  \caption{Results for the topological susceptibility $10^4 t_0^2 \, \chi(r_{\mathrm{cut}})$ 
  using the standard algorithm and the new algorithm that we propose in this paper. 
  The values of $r_{\mathrm{cut}}$ in physical units were computed using the $r_0$ scale 
  from \cite{Necco:2001xg}.}  
  \label{tab:Suscept}
\end{table}

\section{Conclusion}

In this paper we have studied a multi-level algorithm for computing the two point
correlation function of flow observables. It is based on the
idea originally introduced in~\cite{Luscher:2001up}. 
Basically, we split the lattice into two sub-volumes separated by a boundary $B$ and 
use the locality of the action to perform independent updates on each of them. 
Such an approach would not work for observables at positive flow time, so we slightly 
modify the flow equations to build a ``good'' approximation of the original observable 
which can be factorized as required for a multi-level type scheme to 
work. 

In this type of algorithms one starts by performing $N_0$ standard updates 
followed by $N_1$ nested updates for each of the original $N_0$ generated 
configurations. In the ideal case one expects the scaling of the error to be 
proportional to $1/N_1$ instead of the standard $1/\sqrt{N_1}$. We put this 
to the test and for the case of the connected two point correlation function 
$\ev{\mathcal{O}(x)\mathcal{O}(y)} - \ev{\mathcal{O}(x)}\ev{\mathcal{O}(y)}$ 
we find that our algorithm outperforms the standard one when $x$ and $y$ are 
far from the boundary $B$ in units of $1/m_0$ and of the flow radius $\sqrt{8t}$, 
where $m_0$ is the lightest mass compatible with the observable $\mathcal{O}$. 
In the case of short separations our algorithm 
is not better than the standard one, which is expected from the way 
the observables are constructed. 

We also showed that our algorithm can be used to obtain a better lattice 
determination of the topological susceptibility $\chi$, where the 
large statistical errors coming from the tail of the correlator are tamed. 
With our choice of parameters, we observe a decrease of errors by a factor 
larger than two for the same statistics as the standard algorithm, which would 
correspond to a fivefold decrease of the computational time required for a 
fixed target error.

Although we performed our analysis with the Yang-Mills energy density $e$ and the 
topological charge $q$, the idea can be applied to any correlation 
function of flow observables in the lattice Yang-Mills gauge theory. Also, the idea 
that we presented in this paper can be generalized to a four dimensional approach in 
which the decomposition is not limited to the time direction. In that case we expect 
an even better performance of the algorithm.

\begin{acknowledgement}
We are very thankful to R. Sommer for extensive discussions. 
We also would like to thank L. Giusti, M. C\`e and D. Banerjee 
for discussions related to multi-level algorithms. 
Our simulations were performed at the ZIB computer center with the 
computer resources granted by The North-German 
Supercomputing Alliance (HLRN). M.G.V acknowledges the 
support from the Research Training Group GRK1504/2 ``Mass, Spectrum, Symmetry'' 
founded by the German Research Foundation (DFG). 

\end{acknowledgement}


\begin{appendices}
\section{Error reduction}\label{sec:Appendix}

In a two-level nested Monte Carlo algorithm as the one described in the main text, we are 
interested in the scaling of errors with 
respect to $N_0$ and $N_1$. In particular, we look at the case of the two point 
correlator

\begin{equation*}
A = \ev{\mathcal{O}(x_0) \mathcal{O}(y_0)} \, ,
\end{equation*}
where $\mathcal{O}(x_0) \in L$ and $\mathcal{O}(y_0) \in R$. To simplify the notation 
we write $\mathcal{O} \equiv \mathcal{O}(x_0)$ and $\mathcal{O'} \equiv \mathcal{O}(y_0)$.

In a Monte Carlo simulation, an estimator for $A$ is given by

\begin{equation}\label{eq:A_estimator}
\hat{A} = \frac{1}{N_0} \sum_{i=1}^{N_0} \frac{1}{N^2_1} \sum_{j=1}^{N_1} \sum_{k=1}^{N_1} 
\mathcal{O}^{ij} \mathcal{O'}^{ik} \, .
\end{equation}
The error $\sigma^2_A$ on the estimator is then computed in the usual way

\begin{equation}\label{eq:A_error}
\sigma^2_A = \ev{\left(\hat{A} - \bar{A}\right)^2}_{LBR} \, ,
\end{equation}
where $\ev{~}_{LBR}$ stands for the average over all the gauge links in 
$L \cup B \cup R$, and $\bar{A}= \ev{\left[\mathcal{O}\right]_L \left[\mathcal{O'}\right]_R}_B$ 
is the real expectation value of $A$. 

By inserting $\hat{A}$ from Eq.~\eqref{eq:A_estimator} into Eq.~\eqref{eq:A_error} 
and using the fact that the $N_0$ updates are independent one obtains

\begin{multline}\label{eq:A_scaling}
\sigma^2_A = \frac{1}{N_0 N^2_1} \ev{\text{Var}_L\left(\mathcal{O}\right)\text{Var}_R\left(\mathcal{O'}\right)}_B
+ \frac{1}{N_0} \left( \ev{\left[\mathcal{O}\right]^2_L \left[\mathcal{O}'\right]^2_R}_B - \bar{A}^2 \right) +\\
+\frac{1}{N_0 N_1} \left( \ev{\text{Var}_L\left(\mathcal{O}\right) \left[\mathcal{O}'\right]^2_R + \text{Var}_R\left(\mathcal{O}'\right) \left[\mathcal{O}^{~}\right]^2_L}_B   \right)\, ,
\end{multline}
where $\text{Var}_L\left(\mathcal{O}\right) = \left[\mathcal{O}^2\right]_L - \left[\mathcal{O}\right]_L^2$ and 
similarly for $\text{Var}_R \left(\mathcal{O'}\right)$. By looking at Eq.~\eqref{eq:A_scaling} 
it is clear that the error scales not only as the ideal case $1/\sqrt{N_0}N_1$, but it has 
also subleading contributions.

Note however, that using the transfer matrix formalism, one can show that the second 
term proportional to $1/N_0$ is exponentially suppressed as $e^{-m_0 | x^B_0 - x^M_0|}$, 
where $m_0$ is the mass of the lightest state compatible with the 
symmetries of $\mathcal{O}$ and $x^M_0$ corresponds to $x_0$ or $y_0$, whichever 
is the closest to $x^B_0$. 

The third term is also exponentially suppressed if one considers the case of 
the connected correlator 

\begin{equation*}
C=\ev{\mathcal{O}(x_0) \mathcal{O}(y_0)} -\ev{\mathcal{O}(x_0)}\ev{\mathcal{O}(y_0)}\, .
\end{equation*}

Then only the first term gives the leading contribution to the error and it is 
the one that has the ideal scaling for which a nested Monte Carlo scheme would 
be useful. 

The final formula for the error of the connected correlator is

\begin{equation}\label{eq:C_scaling}
\sigma^2_C \approx \frac{1}{N_0 N^2_1} \ev{\text{Var}_L\left(\mathcal{O}\right)\text{Var}_R\left(\mathcal{O'}\right)}_B
+ e^{-m_0 | x^B_0 - x^M_0|} \left( \frac{c_1}{N_0N_1} + \frac{c_2}{N_0}   \right) \, .
\end{equation}

\end{appendices}

\providecommand{\href}[2]{#2}\begingroup\raggedright\endgroup


\begin{thebibliography}{10}

\bibitem{Parisi:1983ae}
G.~Parisi, {\it {The Strategy for Computing the Hadronic Mass Spectrum}},  {\em
  Phys. Rept.} {\bf 103} (1984) 203--211.

\bibitem{Narayanan:2006rf}
R.~Narayanan and H.~Neuberger, {\it {Infinite N phase transitions in continuum
  Wilson loop operators}},  {\em JHEP} {\bf 03} (2006) 064,
  [\href{http://arxiv.org/abs/hep-th/0601210}{{\tt hep-th/0601210}}].

\bibitem{Luscher:2010iy}
M.~L{\"u}scher, {\it {Properties and uses of the Wilson flow in lattice QCD}},
  {\em JHEP} {\bf 08} (2010) 071, [\href{http://arxiv.org/abs/1006.4518}{{\tt
  arXiv:1006.4518}}]. [Erratum: JHEP03,092(2014)].

\bibitem{Luscher:2011bx}
M.~L{\"u}scher and P.~Weisz, {\it {Perturbative analysis of the gradient flow
  in non-abelian gauge theories}},  {\em JHEP} {\bf 02} (2011) 051,
  [\href{http://arxiv.org/abs/1101.0963}{{\tt arXiv:1101.0963}}].

\bibitem{Ce:2015qha}
M.~C{\`e}, C.~Consonni, G.~P. Engel, and L.~Giusti, {\it {Non-Gaussianities in
  the topological charge distribution of the SU(3) Yang--Mills theory}},  {\em
  Phys. Rev.} {\bf D92} (2015), no.~7 074502,
  [\href{http://arxiv.org/abs/1506.0605}{{\tt arXiv:1506.0605}}].

\bibitem{Bazavov:2010xr}
{\bf MILC} Collaboration, A.~Bazavov et~al., {\it {Topological susceptibility
  with the asqtad action}},  {\em Phys. Rev.} {\bf D81} (2010) 114501,
  [\href{http://arxiv.org/abs/1003.5695}{{\tt arXiv:1003.5695}}].

\bibitem{Bruno:2014ova}
{\bf ALPHA} Collaboration, M.~Bruno, S.~Schaefer, and R.~Sommer, {\it
  {Topological susceptibility and the sampling of field space in N$_{f}$ = 2
  lattice QCD simulations}},  {\em JHEP} {\bf 08} (2014) 150,
  [\href{http://arxiv.org/abs/1406.5363}{{\tt arXiv:1406.5363}}].

\bibitem{Chowdhury:2014kfa}
A.~Chowdhury, A.~Harindranath, and J.~Maiti, {\it {Open Boundary Condition,
  Wilson Flow and the Scalar Glueball Mass}},  {\em JHEP} {\bf 06} (2014) 067,
  [\href{http://arxiv.org/abs/1402.7138}{{\tt arXiv:1402.7138}}].

\bibitem{Parisi:1983hm}
G.~Parisi, R.~Petronzio, and F.~Rapuano, {\it {A Measurement of the String
  Tension Near the Continuum Limit}},  {\em Phys. Lett.} {\bf B128} (1983) 418.

\bibitem{Luscher:2001up}
M.~L{\"u}scher and P.~Weisz, {\it {Locality and exponential error reduction in
  numerical lattice gauge theory}},  {\em JHEP} {\bf 09} (2001) 010,
  [\href{http://arxiv.org/abs/hep-lat/0108014}{{\tt hep-lat/0108014}}].

\bibitem{Ce:2016idq}
M.~Cè, L.~Giusti, and S.~Schaefer, {\it {Domain decomposition, multi-level
  integration and exponential noise reduction in lattice QCD}},
  \href{http://arxiv.org/abs/1601.0458}{{\tt arXiv:1601.0458}}.

\bibitem{Meyer:2002cd}
H.~B. Meyer, {\it {Locality and statistical error reduction on correlation
  functions}},  {\em JHEP} {\bf 01} (2003) 048,
  [\href{http://arxiv.org/abs/hep-lat/0209145}{{\tt hep-lat/0209145}}].

\bibitem{Luscher:2009eq}
M.~L{\"u}scher, {\it {Trivializing maps, the Wilson flow and the HMC
  algorithm}},  {\em Commun. Math. Phys.} {\bf 293} (2010) 899--919,
  [\href{http://arxiv.org/abs/0907.5491}{{\tt arXiv:0907.5491}}].

\bibitem{Luscher:2011kk}
M.~L{\"u}scher and S.~Schaefer, {\it {Lattice QCD without topology barriers}},
  {\em JHEP} {\bf 07} (2011) 036, [\href{http://arxiv.org/abs/1105.4749}{{\tt
  arXiv:1105.4749}}].

\bibitem{Necco:2001xg}
S.~Necco and R.~Sommer, {\it {The N(f) = 0 heavy quark potential from short to
  intermediate distances}},  {\em Nucl. Phys.} {\bf B622} (2002) 328--346,
  [\href{http://arxiv.org/abs/hep-lat/0108008}{{\tt hep-lat/0108008}}].

\bibitem{Wolff:2003sm}
{\bf ALPHA} Collaboration, U.~Wolff, {\it {Monte Carlo errors with less
  errors}},  {\em Comput. Phys. Commun.} {\bf 156} (2004) 143--153,
  [\href{http://arxiv.org/abs/hep-lat/0306017}{{\tt hep-lat/0306017}}].
  [Erratum: Comput. Phys. Commun.176,383(2007)].

\bibitem{Cabibbo:1982zn}
N.~Cabibbo and E.~Marinari, {\it {A New Method for Updating SU(N) Matrices in
  Computer Simulations of Gauge Theories}},  {\em Phys. Lett.} {\bf B119}
  (1982) 387--390.

\bibitem{Brown:1987rra}
F.~R. Brown and T.~J. Woch, {\it {Overrelaxed Heat Bath and Metropolis
  Algorithms for Accelerating Pure Gauge Monte Carlo Calculations}},  {\em
  Phys. Rev. Lett.} {\bf 58} (1987) 2394.

\end{thebibliography}
\end{document}